\def\al{\alpha}
\def\be{\beta}
\def\ga{\gamma}
\def\de{\delta}
\def\ep{\epsilon}

\def\ka{\kappa}
\def\la{\lambda}

\def\si{\sigma}

\def\ps{\psi}

\def\Ga{\Gamma}
\def\De{\Delta}

\def\La{\Lambda}

\def\mn{{\mu\nu}}

\def\lrcov{\stackrel{\leftrightarrow}{D}}

\def\half{{\textstyle{1\over 2}}}

\newcommand{\beq}{\begin{equation}}
\newcommand{\eeq}{\end{equation}}
\newcommand{\bea}{\begin{eqnarray}}
\newcommand{\eea}{\end{eqnarray}}
\newcommand{\rf}[1]{(\ref{#1})}

\documentclass{article}  

\begin{document}

\begin{center}
{{\bf New Implications of Lorentz Violation \\}
\vglue 1.0cm
{Don Colladay\\} 
\bigskip
{\it New College of Florida\\}
\medskip
{\it Sarasota, FL, 34243, U.S.A.\\}
 }
\end{center}
\vglue 0.8cm

\vglue 0.3cm
 
{\rightskip=3pc\leftskip=3pc\noindent
In this proceedings, I summarize two recently discovered 
theoretical implications that Lorentz violation has on
physical systems.  First, I discuss new models for neutrino 
oscillations in which relatively simple combinations of
Lorentz-violating parameters can mimic the major features
of the current neutrino oscillation data. 
Second, I will present results on Yang-Mills instantons
in Lorentz-violating background fields.  
An explicit solution is presented for unit winding
number in $SU(2)$.
}

\vskip 1 cm

\vskip 1 cm

\section{Introduction}

Enormous success in particle physics has been obtained during the last century by 
assuming symmetry of the fundamental action under the Lorentz group.
Supplementing this with various assumed gauge symmetries and
representational content eventually led to the standard model.
A key step in constructing the standard model involves spontaneously breaking
one of these assumed symmetries as well as relaxing some of the 
discrete symmetries in the electroweak sector.
A natural question arises as to the validity of perfect symmetry
under the Lorentz group as well.

In fact, there are theoretical reasons to suspect that Lorentz symmetry breaking
may arise naturally in more fundamental theories such as string theory\cite{kps}
or other attempts at quantum gravity\cite{qgrav}.
In addition, there are numerous experimental tests of Lorentz invariance
in a variety of sectors\cite{cptproc}.
A general framework for including general Lorentz breaking effects into 
the standard model
has been constructed\cite{kp,ck}.  The resulting effective field theory is called the
Standard Model Extension (SME).
Stability and causality issues as well as generic properties of the dispersion
relations have also been studied\cite{lehnert}.

\section{Lorentz and CPT Violation}
For about the past fifteen years, it has been known that miniscule remnant
effects that violate Lorentz invariance may arise in a more fundamental
theory of nature\cite{kps}.
In addition, the well known CPT theorem proves that any local, 
Lorentz invariant quantum field theory must also preserve CPT.
In fact, this theorem has been expanded to prove that CPT violation
implies Lorentz violation\cite{green}, demonstrating that bounds on
CPT can be interpreted as bounds on Lorentz violation.

The generic features of such violations may be incorporated into effective
field theory using a generic spontaneous symmetry breaking mechanism that 
is analogous to the conventional Higg's mechanism of the standard model.
The crucial difference concerns the field that exhibits a nonzero vacuum
expectation value.
In conventional Higg's models, the field that is used to break electroweak 
symmetry is taken as a scalar field in order to preserve Lorentz invariance
as well as renormalizability.
Consider a generic field theory containing gauge bosons with tensor indices 
($B^\mu$ for example) with nontrivial couplings to the fermions
(terms of the type $B^\mu \overline \psi \gamma^5 \gamma_\mu \psi$ for example).
A Lorentz covariant potential for the tensor field can induce a nonzero
expectation value of the form $\langle{B^\mu}\rangle$ that will generate
Lorentz-violating contributions to the matter sectors.

The SME consists of all possible terms that couple the standard model fields
to background tensor fields.\cite{kp,ck}
It is the spirit of the model to be as general as possible so that any 
experiment that exhibits Lorentz violation in the future can be described
in this formalism.  
The hope is that experimentally identifying specific constants for Lorentz 
violation that occur in nature may serve as a window to a more fundamental 
theory.
On the theoretical side, the SME is general enough to accommodate
any theory that involves Lorentz Violation. 
For example, it has been argued that any realistic theory of 
noncommutative geometry must reduce to a subset of the SME\cite{kl}.
For practical calculations, it is often useful to restrict the
couplings to a minimal set that preserves the conventional gauge
invariance of the standard model as well as power counting
renormalizability.  Imposing translational invariance on these
couplings yields the minimal SME, useful for quantifying leading
order corrections to experiments.

As an example, consider the electron-photon sector.
Imposing gauge invariance and restricting to power-counting
renormalizable terms in the standard model extension yields a lagrangian of 
\begin{equation}
L =
\half i \overline{\ps}  \Ga^\nu \lrcov_\nu \ps 
- \overline{\ps}  M  \ps
+ L_{\rm photon}
\quad ,
\label{baselag}
\end{equation}
where $\Ga$ and $M$ denote
\beq
{\Ga}^{\nu}  ={\ga}^{\nu}+  c^{\mu \nu} 
{\ga}_{\mu}+  d^{\mu \nu} {\ga}_{5} {\ga}_{\mu}
\quad ,
\eeq
\beq
 M  = m+  a_{\mu}  {\ga}^{\mu}+  b_{\mu}
 {\ga}_{5} {\ga}^{\mu}+\frac{1}{2} H^{\mu \nu}
 {\sigma}_{\mu \nu}
\quad .
\eeq 
The parameters $a$, $b$, $c$, $d$, and $H$ are related to fixed background
expectation values of tensor fields.
In this sector, stringent bounds on many parameters
have been attained.
For example, limits on the order $|k|<10^{-32}$ for photons\cite{km2},
and $|b_3|< 10^{-24}m_e$ for electrons\cite{mhs} have been obtained.

Different sectors of the SME have independent parameters for the 
background fields, therefore the stringent limits in electrodynamics
do not rule out potentially large effects in other sectors.
For example, as I will discuss next, it may be possible that current
experimental data regarding neutrino oscillations can be modeled using
Lorentz violating terms, rather than masses.

\section{Application to Neutrino Oscillations}
The conventional formalism appears to describe much of the current data
involving neutrino masses fairly well using mass differences on the
order of $\De m^2 \sim 10^{-20}GeV^2$.  
The ratio $\De m^2 / E^2 \sim 10^{-20}$ happens to be compatible
with leading order Planck suppression estimates of the Lorentz-Violation parameters.
It is therefore reasonable to ask if these oscillation effects are
really a manifestation of Lorentz-violating background fields coupled
to neutrinos.

The SME effective hamiltonian for neutrinos and antineutrinos in the 
presence of Lorentz violation has recently been constructed\cite{ck,km}.
This model is important as it includes all possible leading order
corrections to the neutrino propagators in the presence of Lorentz 
violation.
All previous work on neutrinos in the presence of Lorentz violation has
assumed a rotationally invariant subset of the SME
(called Fried Chicken models)
typically with two neutrino species and nonzero
neutrino masses\cite{cgn}.
The general case that includes three neutrino species
and allows for violation of rotational symmetry 
can be expressed using an effective hamiltonian
in the active neutrino basis $(\nu_a, \overline \nu_a)$,
where $a$ represents $e$, $\mu$, or $\tau$.
\beq
(h_{eff})_{ab} =  
|\vec p|\de_{ab} 
\left(\begin{array}{cc} 
1 & 0 \\ 
0 & 1 
\end{array}\right) 
+\frac{1}{2|\vec p|} 
\left(\begin{array}{cc} 
(\tilde m)_{ab} & 0 \\ 
0& (\tilde m)^*_{ab} 
\end{array}\right) 
\quad+\frac{1}{|\vec p|} 
\left(\begin{array}{cc} 
M_{11} & 
M_{12} \\ 
M_{21} & 
M_{22}
\end{array}\right) 
, 
\eeq 
where
\beq
M_{11} = [(a_L)^\mu p_\mu-(c_L)^\mn p_\mu p_\nu]_{ab}
\quad ,
\eeq
\beq
M_{12} = -i\sqrt{2} p_\mu (\ep_+)_\nu 
[(g^{\mn\si}p_\si-H^\mn){C}]_{ab}
\quad ,
\eeq
\beq
M_{21} = i\sqrt{2} p_\mu (\ep_+)^*_\nu 
[(g^{\mn\si}p_\si+H^\mn){C}]^*_{ab}
\quad ,
\eeq
\beq
M_{22} = [-(a_L)^\mu p_\mu-(c_L)^\mn p_\mu p_\nu]^*_{ab}
\quad .
\eeq
For illustrative purposes, this form can be restricted to the minimal SME\cite{ck}
for which only the left-handed neutrino doublet $L_a$
is present.
The resulting lagrangian contains the terms
\beq
L \supset \frac i 2 \overline L_a \ga^\mu 
\stackrel{\leftrightarrow}{D_\mu} L_a
 - (a_L)_{\mu a b} \overline L_a \ga^\mu L_b
 + \frac i 2 (c_L)_{\mn a b} \overline L_a \ga^\mu 
 \stackrel{\leftrightarrow}{D_\mu} L_b
 \quad ,
\eeq
yielding the effective neutrino hamiltonian
\beq
(h_{eff})_{ab} = |\vec p | \de_{ab}
+ {1 \over |\vec p|} [(a_L)^\mu 
p_\mu - (c_L)^\mn   
p_\mu p_\nu]_{ab}
\quad .
\eeq
Note that $c_L$ ($a_L$) preserves (violates) CPT.
Diagonalization of this matrix yields two momentum
dependent eigenvalue differences that govern
the neutrino and antineutrino oscillation probabilities.

Some generic features of these oscillation probabilities
may be identified by analyzing dimensionless combinations of parameters
that appear in the oscillatory function arguments.
For the standard massive neutrino case, the relevant ratio
is $\De m^2 \cdot (L/E)$.
The Lorentz violation terms typically contribute $a_L \cdot (L)$ and
$c_L \cdot (LE)$ factors in the argument.
This means that novel new energy dependences for the oscillations
may be attained.
In general, there will also be rotationally noninvariant terms
contributing to the oscillation arguments.
This opens up the possibility for interesting searches for
diurnal variations at the Earth's siderial period $\omega \approx 23~ h~ 56 m$.
A realistic model within the minimal SME that appears consistent 
with current experimental data is the bicycle model\cite{km}.
This model is notable since it consists of a two parameter fit 
to the currently observed data, while at the same time maintaining
the full gauge invariance of the standard model.
The bicycle model sets all Lorentz violating parameters
to zero, except the rotationally invariant piece of
$c_L$ and a single spatial componet of $a_L$.

Regardless of the specific choice of parameters, there
are specific signatures for Lorentz Violation in neutrino
oscillations.  They are:
\begin{itemize}
\item{Spectral anomalies (L or L/E oscillation behavior).}
\item{L - E conflicts for experiments in different regions of L - E space
that cannot be accommodated using only two mass differences.}
\item{Periodic Variations, such as a diurnal signal.}
\item{Compass asymmetries (effects that cannot be attributed 
standard physics such as the
effect of the Earth's magnetic field on cosmic rays).}
\item{Neutrino-antineutrino mixing.}
\item{Classic CPT test: $P_{{\nu_b}\rightarrow \nu_a} \ne 
P_{{\overline \nu_a}\rightarrow \overline \nu_b}$.}
\end{itemize}
Note that the only one of these for which there is a
possible signal is the L - E conflict of LSND\cite{lsnd} to be
tested by the future data collected by MiniBooNE.

\section{Yang-Mills Instantons with Lorentz Violation}
Static solutions to pure Yang-Mills theories in four Euclidean 
dimensions are well known and are called instantons.
The pure Yang-Mills sector of the SME contains terms that violate the Lorentz symmetry,
but it turns out that many of the properties of instanton solutions
remain intact\cite{cm}.
This result is due to the fact that the instanton solutions
rely heavily upon topological arguments as will be discussed
in the remainder of this proceedings.

The standard pure Yang-Mills Euclidean action is given by
\beq
S_0(A) = {1 \over 2} \int d^4 x ~ Tr[F^\mn F^\mn]
\quad ,
\eeq
where
\beq
F^\mn = \partial^\mu A^\nu - \partial^\nu A^\mu + i g [A^\mu, A^\nu] \quad ,
\eeq
is the curvature of the connection $A$.
The topological charge $q$ is defined
as
\beq
q = {g^2 \over 16 \pi^2} \int d^4 x Tr \tilde F^\mn F^\mn
\quad ,
\eeq
where 
$\tilde F^\mn = \frac 1 2 \ep^{\mn\al\be} F^{\al\be}$ 
is the dual of $F$.
A useful identity is:
$\frac 1 4 Tr \tilde F F = \partial^\mu X^\mu$,
with
\beq
X^\mu = \frac 1 4 \ep^{\mn\la\ka}Tr(A^\nu F^{\la\ka} - \frac 2 3 i g A^\nu A^\la A^\ka)
\quad .
\eeq
This converts the topological charge integral to a surface integral
with the net result that $q$ must be an integer.
Note that this argument is independent of the specific form of the action.

The equation of motion for the curvature is 
\beq
[D^\mu, F^\mn] = 0
\quad ,
\eeq
with a corresponding Bianchi Identity that 
follows from the definition of $F$:
\beq
[D^\mu, \tilde F^\mn] = 0
\quad .
\eeq
($D_\mu = \partial_\mu + i q A_\mu$ is the usual covariant
derivative)
This gives a set of nonlinear differential equations for $A^\mu$.
A clever argument for solving\cite{bpst} these equations involves
consideration of the inequality
\beq
\frac 1 2 \int d^4 x Tr(F \mp \tilde F)^2 \ge 0
\quad .
\eeq
This can be rearranged as
\beq
S \ge \pm \frac 1 2 \int d^4 x Tr[\tilde F^\mn F^\mn] = \pm {8 \pi^2 \over g^2} q 
\quad .
\eeq
The inequality is saturated for $F = \pm \tilde F$, implying that
self-dual or anti-self-dual curvatures are extremal solutions.

As an example of an explicit self-dual solution, 
let $q = 1$ , with gauge group G = SU(2).
The vector potential can be written as
\beq
A^\mu_{SD} = - {\tau^\mn x^\nu \over g(\rho^2 + x^2)}
\quad ,
\eeq
and the corresponding curvature is
\beq
F^\mn_{SD} = {2 \rho^2 \over g(\rho^2 + x^2)^2} \tau^\mn
\quad ,
\eeq
where $\tau^{0i} = \si^i$ 
and $\tau^{ij} = \ep^{ijk} \si^k$ are written in terms
of the conventional Pauli sigma matrices.
The free parameter $\rho$ controls the instanton size.
The anti-self-dual solution ($q = -1$) is obtained using the 
parity transform of the above solution.
Subsequently, all self-dual solutions were classified\cite{adhm}.

Next, the Lorentz-violating case is examined.
The quadratic action that preserves gauge invariance is given
by
\beq
S(A) = {1 \over 2} \int d^4x Tr[(F^\mn F^\mn) + (k_F)^{\mn \alpha \beta} 
F^\mn F^{\al \be}]
\quad ,
\eeq 
where the parameters $k_F$ 
are small, constant background fields. 
Only terms of $O(k_F)$ are kept in the calculations.
The first result is that the topological charge $q$
remains integral because the conventional argument
is insensitive to the detailed form of the action, provided
that gauge invariance is maintained.
A modified bound on the action is
\beq
S \ge \pm {8 \pi^2 \over g^2} q \pm \frac 1 4 
(k_F^{\mn\al\be} + \tilde k_F^{\mn \al \be})\int d^4 x Tr
\tilde F^\mn F^{\al\be}
\quad ,
\label{act}
\eeq
where $\tilde k_F^{\mn\al\be} 
= \frac 1 4 
\ep^{\mn\la\ka} k_F^{\la\ka\rho\si} 
\ep^{\rho\si\al\be}$.
It is useful to decompose $k_F = k_{F}^+ \oplus k_{F}^-$ 
according to its duality properties
and consider the two cases separately.

For case one ($k_{F} = - \tilde k_{F})$, the background constants 
take the form $k_F^{\mn\al\be} = \La^{[\mu[\al} \de^{\nu]\be]}$,
where $\La^\mn$ is a symmetric, traceless matrix. 
The action is then extremal for the modified duality condition 
\beq
F^\prime \simeq \pm \tilde F^\prime
\quad ,
\eeq
where
$F^{\prime \mn} = F^\mn + \frac 1 2 k_F^{\mn\al\be} F^{\al\be}$.
Explicit solutions are constructed using 
$\tilde x^\mu = x^\mu + \La^\mn x^\nu$,
and the vector potential is given by
$A^\mu(x) \simeq A^\mu_{SD}(\tilde x) + \La^{\mn} A^\nu_{SD}(x)$.
These solutions take the form of conventional instantons in skewed coordinates.

For case two ($k_{F} = \tilde k_{F}$), the background constants
are trace free.
In this case, the lower bound on $S$ given by \rf{act} varies
with $\delta F$.
This means that the previous modified duality condition
fails to generate a solution and the equations of motion
must be solved explicitly.
This can be done to leading order in $k_F$ by expanding 
$A = A_{SD} + A_k$, fixing 
$A$ to be close to the conventional self-dual solution.
The equations of motion become
\beq
[D_{SD}^\nu,[D_{SD}^\nu, A_k^\mu]] 
+ 2 i g [F_{SD}^\mn, A_k^\nu] 
= j_k^\mu
\quad ,
\eeq
where $j_k^\mu \equiv k_F^{\mn\al\be}
[D^\nu_{SD}, F^{\al\be}_{SD}]$.

This is a second-order, linear elliptic differential equation.
A formal solution can be constructed using 
the relevant propagator $G(x,y)$:
\beq
A_k = \int d^4 y G(x,y) j_k(y)
\eeq

For the case $q=1$ with G = SU(2), an explicit solution 
can be constructed
using the following procedure:
\begin{itemize}
\item{Transform to singular gauge
$\rightarrow$ makes fields $\sim O(\rho^2)$.}
\item{To $O(\rho^2)$ in this gauge, can use free propagator 
$$G_0(x,y) = {1 \over 4 \pi^2 (x-y)^2} \quad .$$}
\item{Gives tensorial structure for general ansatz
$$A_k^\mu = {2 \rho^2 x^2 \over 3 g} f(x^2) 
k_F^{\mn\al\be}\tau^{\al(\nu} x^{\be)} \quad .$$}
\item{Substitute into the full equation of motion with general $\rho$.}
\end{itemize}
Remarkably this gives a differential equation for $f(x)$,
indicating that the tensorial structure is in fact correct to
all orders in $\rho^2$.

\section{Conclusion}
A general formalism allowing for Lorentz violation 
(and possible resulting CPT violation)in the neutrino 
sector has been developed.
Possible signals for Lorentz violation include anomalous
energy dependence as well as siderial variations.
To date, only a tiny subset of the neutrino 
sector implications have been explored.
In addition, it has been shown that instantons
can still be classified according to the conventional
topological charge.

%
%
%
%

\end{document}